\documentclass[sigconf]{acmart}
\usepackage{colortbl} %taps OK
\usepackage{soul} %taps OK
\usepackage{xcolor} %taps OK
\usepackage{multirow} %taps OK

\usepackage{mathtools} %taps OK

\usepackage{subfigure}   %taps OK

\usepackage{wrapfig} %taps OK

\usepackage{balance} %taps OK

\def\BibTeX{{\rm B\kern-.05em{\sc i\kern-.025em b}\kern-.08emT\kern-.1667em\lower.7ex\hbox{E}\kern-.125emX}}
    
\copyrightyear{2023}
\acmYear{2023}
\setcopyright{acmlicensed}\acmConference[UbiComp/ISWC '23 Adjunct ]{Adjunct Proceedings of the 2023 ACM International Joint Conference on Pervasive and Ubiquitous Computing \& the 2023 ACM International Symposium on Wearable Computing}{October 8--12, 2023}{Cancun, Quintana Roo, Mexico}
\acmBooktitle{Adjunct Proceedings of the 2023 ACM International Joint Conference on Pervasive and Ubiquitous Computing \& the 2023 ACM International Symposium on Wearable Computing (UbiComp/ISWC '23 Adjunct ), October 8--12, 2023, Cancun, Quintana Roo, Mexico}
\acmPrice{15.00}
\acmDOI{10.1145/3594739.3610744}
\acmISBN{979-8-4007-0200-6/23/10}

\begin{document}
\def \sys {\textit{SolStep}}

\title{Eco-Friendly Sensing for Human Activity Recognition }

 \author{Kaede  Shintani}
\email{k-shintani@ist.osaka-u.ac.jp}
\affiliation{%
  \institution{Osaka University}
  \city{Osaka}
  \country{Japan}}
  
\author{Hamada Rizk}
\email{hamada_rizk@f-eng.tanta.edu.eg}
\affiliation{% 
\institution{Osaka University, Osaka, Japan}
\institution{Tanta University, Tanta, Egypt}}

\author{Hirozumi~Yamaguchi}
\email{h-yamagu@ist.osaka-u.ac.jp}
\affiliation{%
  \institution{Osaka University}
  \city{Osaka}
  \country{Japan}}
\renewcommand{\shortauthors}{Kaede Shintani, Hamada Rizk, \& Hirozumi Yamaguchi}

\begin{abstract} 
With the increasing number of IoT devices, there is a growing demand for energy-free sensors. Human activity recognition holds immense value in numerous daily healthcare applications. However, the majority of current  sensing modalities consume energy, thus limiting their sustainable adoption.
In this paper, we present a novel activity recognition system that not only operates without requiring energy for sensing but also harvests energy. Our proposed system utilizes photovoltaic cells,  attached to the wrist and shoes, as eco-friendly sensing devices for activity recognition. By capturing photovoltaic readings and employing a deep transformer model with powerful learning capabilities, the system effectively recognizes user activities.
To ensure robust performance across various subjects, time periods, and lighting conditions, the system incorporates feature extraction and different processing modules.  The evaluation of the proposed system on realistic indoor and outdoor environments demonstrated its ability to recognize activities with an accuracy of 91.7\%.
\color{black}
\end{abstract}


\keywords{
\vspace{-1mm} activity recognition, Energy harvesting, energy-free sensing}

\maketitle

\vspace{-1mm}
\section{Introduction}
With the increase in human activity tracking, monitoring, and analysis, wearable sensors have become a popular choice in recent years. One of the major challenges faced in this domain is powering these devices without frequent battery replacements~\cite{wei2020solarslam}. 

Existing solutions for wearable activity recognition primarily rely on battery-powered sensors, such as accelerometers, gyroscopes, and magnetometers~\cite{9777982,rizk12smartwatch}. These sensors measure various parameters such as acceleration, rotation, and orientation to infer the user's activity. However, frequent battery replacements can be inconvenient for users, and the disposal of used batteries can lead to environmental hazards.
One possible solution is to use solar (photovoltaic) cells as an energy source for these wearable sensors. 

In this paper, we propose a novel approach to use solar cell as energy-free sensor for recognizing human  activities.
The proposed approach utilizes the photocurrent captured by the solar cells to recognize human activities using an efficient learning model. However, using solar cells as sensors presents technical challenges due to the fluctuation in the amount of light received by the cells. Additionally, the photocurrent generated by the cells is generally low, which can make it challenging to recognize the activity patterns accurately.
To overcome these challenges, we propose an efficient learning model based on a transformer architecture \cite{NIPS2017_3f5ee243}. The transformer model has a self-attention mechanism that can capture the temporal relationships between the activity patterns. This approach can reduce the noise caused by the fluctuation in the amount of light received by the solar cells and improve the accuracy of activity recognition.
Moreover, the use of a transformer model architecture enables the possibility of performing online learning on the device, eliminating the need for external computation. This approach can also significantly reduce the energy consumption required for activity recognition.

The proposed system underwent rigorous testing on six distinct participants, encompassing realistic indoor and outdoor environments. The obtained results demonstrate the system's promise, achieving an accuracy of 91.7\% in recognizing human activities. The evaluation highlights the system's robustness and adaptability to previously unseen environments, as it only incurred a minimal 15\% decrease in accuracy. Furthermore, the system exhibited consistent performance across different users, further attesting to its reliability.
In addition to its remarkable performance, this solution presents itself as a sustainable and eco-friendly sensing alternative when compared to existing solutions and sensing technologies available in the market. By leveraging solar cells as sensors, this system eliminates the necessity for frequent battery replacements, consequently reducing the generation of electronic waste. 

\vspace{-2mm}
\section{Related Work}
Human activity recognition (HAR) has been an active area of research with numerous techniques proposed to accurately classify and identify human activities. Traditional approaches in HAR often rely on inertial sensors, such as accelerometers and gyroscopes, to capture motion-related data for activity recognition. These methods utilize various machine learning and deep learning algorithms to analyze and classify activity patterns based on sensor data  \cite{s22249891}.

However, recent advancements have introduced the integration of energy harvesters as context sensors in HAR systems. Energy harvesters, such as Kinetic Energy Harvesting (KEH) transducers are increasingly being used as power-saving activity sensors. Recent studies, such as \cite{ref8}, propose utilizing the energy collected by KEH transducers for HAR, as it offers reduced energy consumption compared to traditional inertial sensors. This feature enables the uninterrupted functioning of IoT devices. Research by Lan et al. \cite{ref26} demonstrates the feasibility of utilizing the voltage signal from a capacitor storing harvested kinetic energy for activity identification.
% KEH transducers find various applications, including monitoring food consumption \cite{ref27}, identifying transportation methods \cite{ref28}, and creating security keys for wearable devices \cite{ref29}. To enhance energy generation through KEH, Ma et al. \cite{ref17} implemented dual transducers in footwear to identify gaits using the harvested signal. Similarly, 
Sandhu et al. \cite{ref14} employed a single KEH transducer for simultaneous energy collection and daily activity recognition. However, the energy harvested from human motion/vibrations alone is insufficient to sustain the operation of wearable devices \cite{ref14}.

Photovoltaic/Solar Energy Harvesting (SEH) offers considerable advantages in power density, energy conversion efficiency, and durability compared to KEH \cite{ref30}. As a result, SEH emerges as an appealing power source for IoT sensor nodes \cite{ref30}. Solar cells have also found use in identifying hand gestures and room-level localization. Research by Ma et al. \cite{ref10,ref11} employ a solar cell to detect and identify various hand gestures in an indoor setting under lamp light. The resulting energy harvesting signal, which contains information about the performed gestures, is due to the hand movement blocking light from reaching the solar surface. Umetsu et al. \cite{ref12} have utilized multisource energy harvesters, including solar and kinetic, for recognizing locations in indoor environments. Nevertheless, systems proposed in \cite{ref10,ref11,ref12} utilized SEH as a context sensor without simultaneous energy extraction, hence not exploiting the full potential of the energy harvesters. However, solar cells are utilized as energy harvesters and context sensors for location tracking in~\cite{9767256,soLoc} and for step counting in~\cite{step}.
Moreover, the application of solar cells for identifying human physical activities in HAR scenarios remains largely unexplored. A preliminary study was proposed in \cite{ref15}.

\textit{
This paper presents a comprehensive study on human activity recognition using solar cells as both an energy harvester and a context sensor. The study investigates the effects of various lighting conditions, timing, and different participants on recognition performance.}

\begin{figure}[!tbp]
\centering \vspace{-0.35cm}
\includegraphics[width=\linewidth,height=2.9cm]{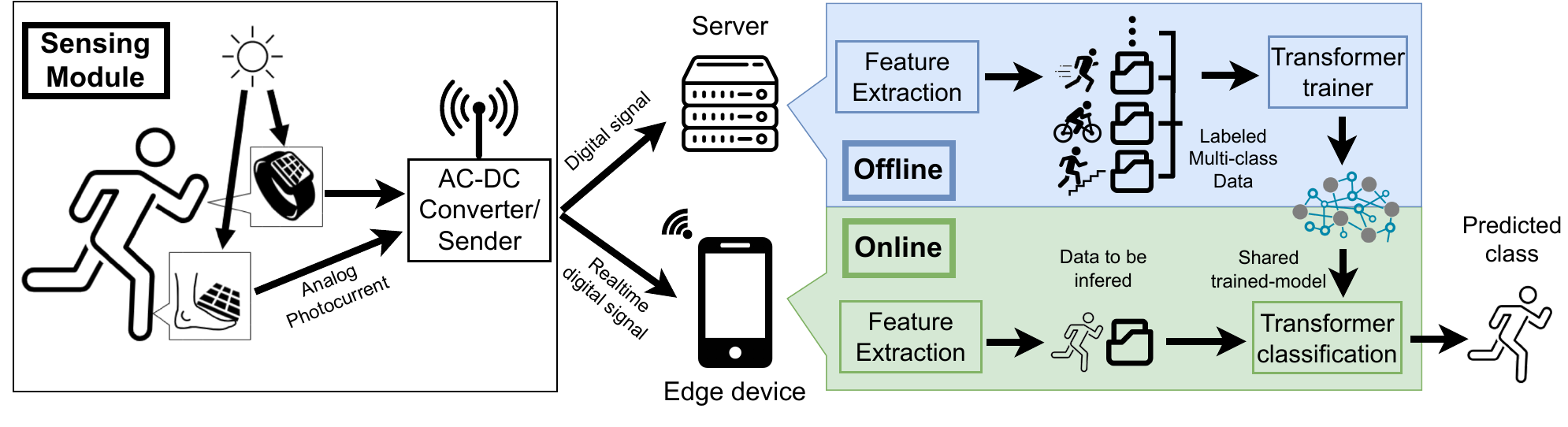} \vspace{-0.9cm}
\caption{System architecture.}
\label{fig:system}
\vspace{-0.55cm} 
\end{figure}
\vspace{-2mm}

\section{System Overview}
\begin{figure}[!tbp]
\centering \vspace{-0.55cm}
\includegraphics[width=0.94\linewidth,height=2.96cm]{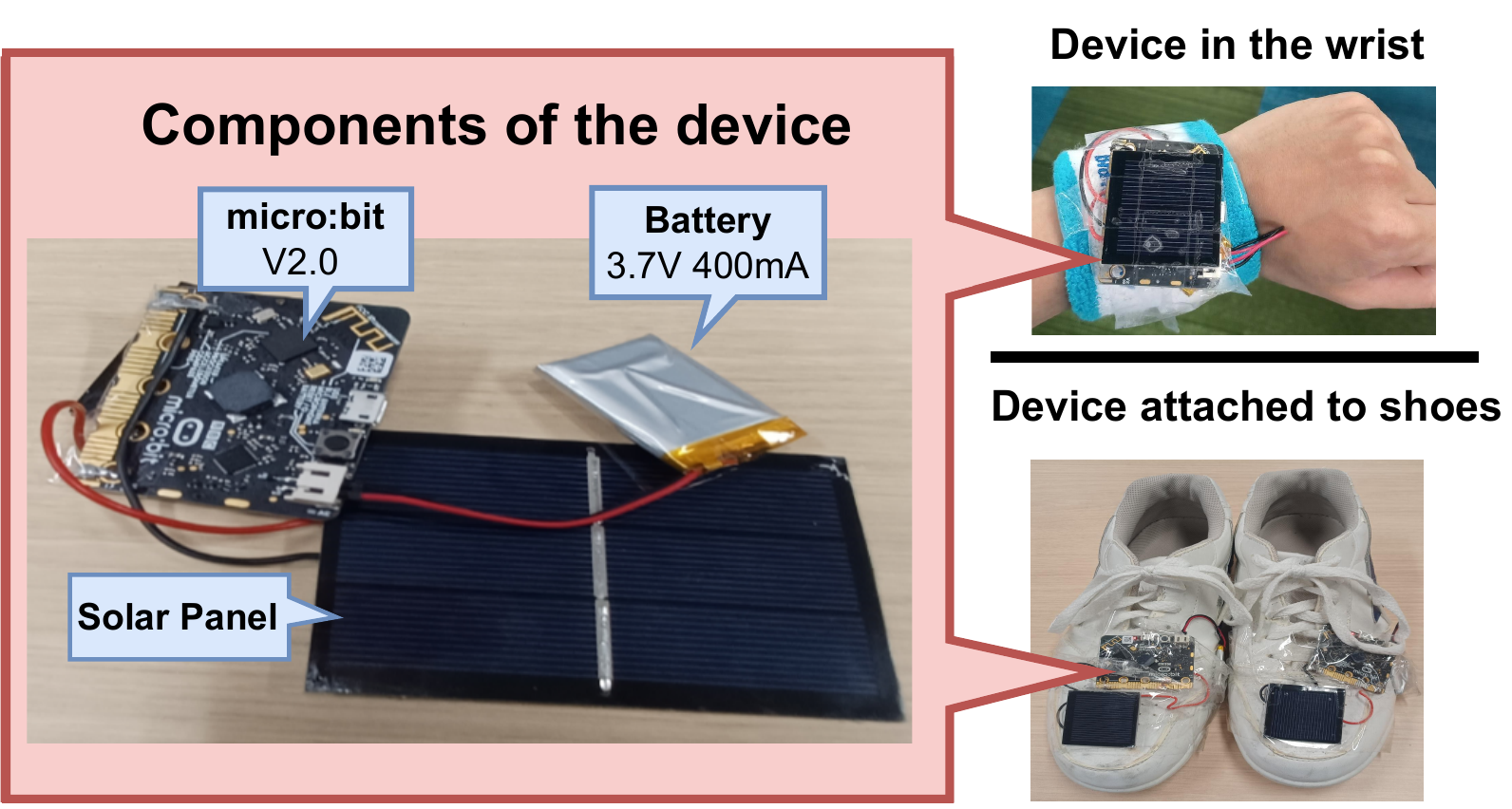} \vspace{-0.35cm}
\caption{Sensing devices.}
\label{fig:real_setup}
\vspace{-0.54cm} 
\end{figure}

The proposed system architecture, as depicted in Fig.~\ref{fig:system}, comprises two phases: an offline phase and an online phase. The offline phase initiates with the sensing and data collection module. This module uses a proprietary compact-sized sensing device developed by our team is securely attached to the participant's shoes and/or wrist, as depicted in Fig. \ref{fig:real_setup}. This device, designed specifically for our system, plays a crucial role in capturing the necessary photocurrent measurements during the participants' activities.
The acquired data, along with the corresponding activity labels, are transmitted to our cloud-based service to construct a fingerprint database for training a recognition model. Subsequently, the preprocessing module is employed to reduce noise in the data, ensuring accurate measurements, and prepare the data in the required format for training the deep recognition model.
Following the preprocessing step, the feature extraction module maps the absolute measurements into relative features, allowing the system to function effectively in diverse lighting conditions and environments. These features are then utilized to train a transformer-based recognition model, capitalizing on the learning capabilities of the transformer model. Once the training phase is complete, the model is deployed on the user's mobile phone for use in the online phase, enabling efficient activity recognition.

During the online phase, photovoltaic measurements are captured by our wearable device and transmitted to the mobile phone via Bluetooth for further sequence preparation and feature extraction. Subsequently, these features are employed to query the recognition model, providing an estimation of the activity being performed.

\color{black}

\vspace{-1mm}
\section{System architecture}
In this section, we describe the proposed approach in detail, including the structure of the transformer model and the processing of the photocurrent input from solar cells to recognize human activities.

\begin{figure}[htbp]
 \vspace{-0.3cm}
    \begin{tabular}{cc}
      \begin{minipage}[t]{0.45\hsize}
        %\centering
        \includegraphics[keepaspectratio, scale=0.3]{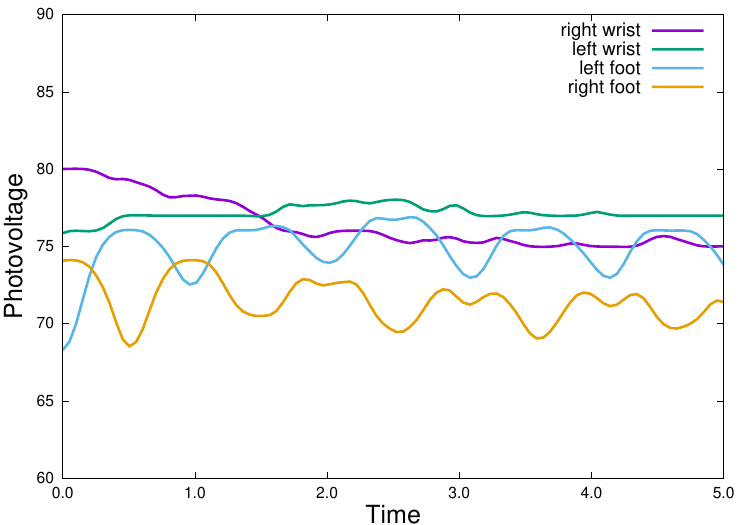}
         \vspace*{-8mm}
        \caption{cycling}
        \label{cycling}
      \end{minipage} &
      \begin{minipage}[t]{0.45\hsize}
        %\centering
        \includegraphics[keepaspectratio, scale=0.3]{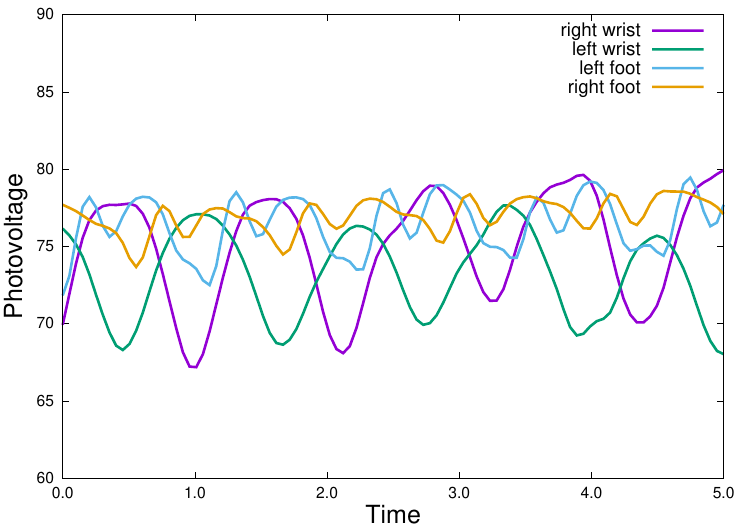}
        \vspace*{-8mm}
        \caption{walking}
        \label{walking}
      \end{minipage} \\
   
      \begin{minipage}[t]{0.45\hsize}
        %\centering
        \includegraphics[keepaspectratio, scale=0.3]{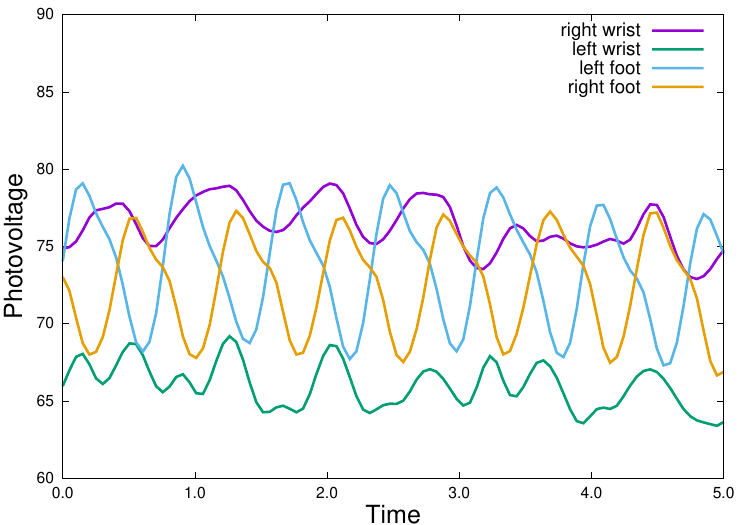}
        \vspace*{-8mm}
        \caption{jogging}
        \label{jogging}
      \end{minipage} &
      \begin{minipage}[t]{0.45\hsize}
        %\centering
        \includegraphics[keepaspectratio, scale=0.3]{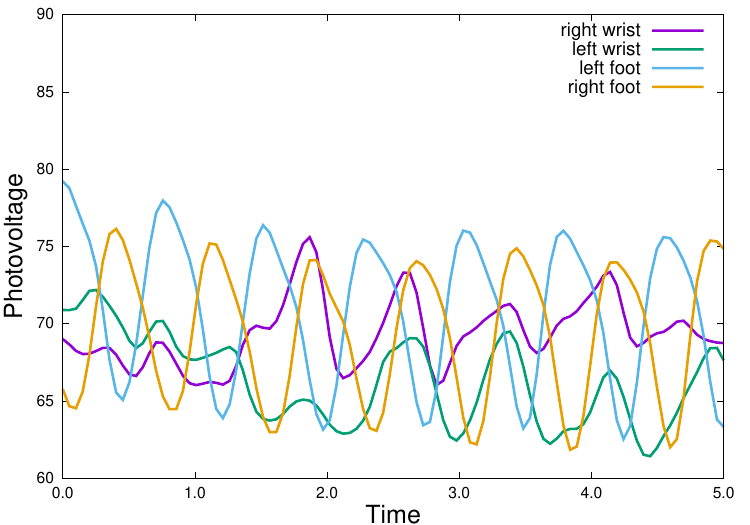}
        \vspace*{-8mm}
        \caption{running}
        \label{running}
      \end{minipage} \\
      \begin{minipage}[t]{0.45\hsize}
       % \centering
        \includegraphics[keepaspectratio, scale=0.3]{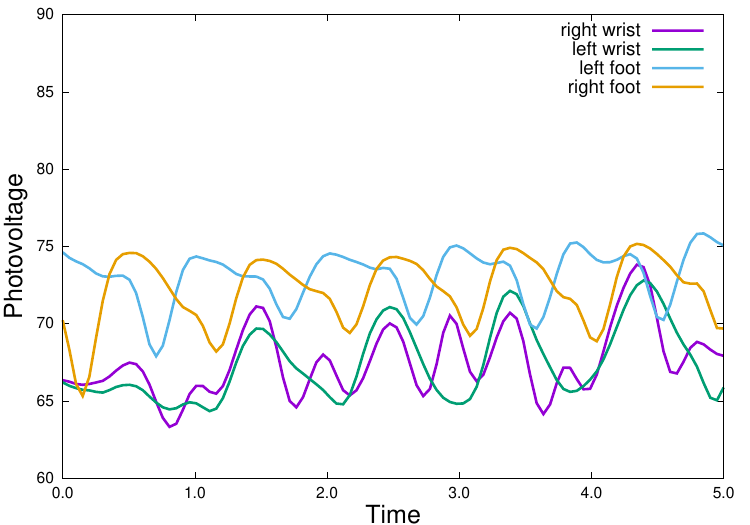}
        \vspace*{-8mm}
        \caption{up}
        \label{up}
      \end{minipage} &
      \begin{minipage}[t]{0.45\hsize}
        \centering
        \includegraphics[keepaspectratio, scale=0.3]{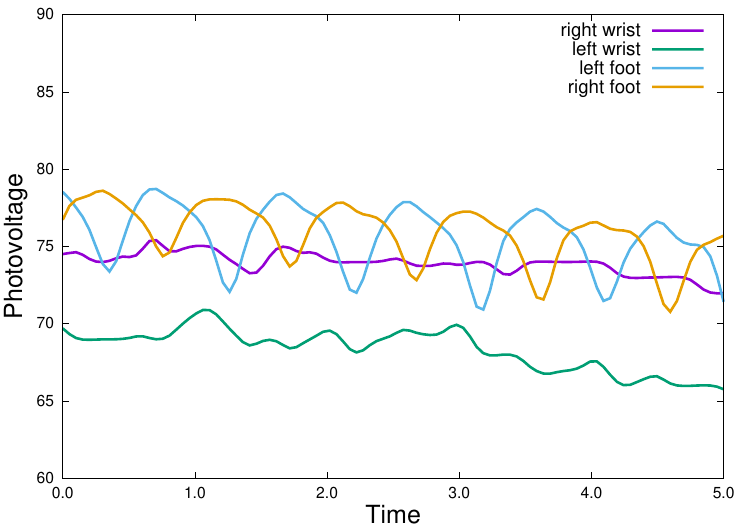}
        \vspace*{-8mm}
        \caption{down}
        \label{down}
      \end{minipage} \\      
    \end{tabular}
    % \caption{hoge}
\vspace{-0.6cm}
  \end{figure}

\subsection{Sensing Module}
Sensing module of the proposed system plays a crucial role in capturing the time-series data required for recognizing human  activities. It comprises three main components: the solar cell, the electronic circuit, and the computing unit.
The solar cell used in the sensing module is a commodity photovoltaic cell with dimensions of 8cm in length, 6cm in width, and 0.1cm in depth. The solar cell has a high efficiency of 15.5\%, which makes it an excellent candidate for capturing the necessary energy from ambient light to power the sensing module.
To optimize energy usage during sensor idle time, the harvested energy is directed towards charging a small attached battery. This approach also stabilizes the operation of the computing and communication modules in low-light scenarios.
To obtain accurate photocurrent measurements from the solar cell, it is connected to a 10-bit analog-to-digital converter (ADC) built into micro:bit. The ADC is responsible for converting the analog photocurrent signal into a digital signal that can be processed by the computing unit. The ADC is then connected to a low-energy-compute and transmission unit, specifically a micro:bit, through a Serial Peripheral Interface, as shown in Figure~\ref{fig:real_setup}.
The measurements are obtained using a Python implementation that sends an HTTP request to the micro:bit to obtain the photocurrent reading response from the solar cell. The Python implementation provides a simple and efficient way to interact with the micro:bit and collect the required data. 
This module is mounted to the smart shoes and wrist of the user, allowing for unobtrusive and seamless monitoring of the user's physical activities.

The photocurrent readings obtained from the solar cell are then converted into voltage to enhance interpretability. The voltage measurements corresponding to six different activities, captured by four sensors placed on the wrist and shoes, are visually represented in Figures~\ref{cycling} to \ref{down}. These distinct photovoltaic patterns provide compelling evidence for the effectiveness of photovoltaic cells as efficient sensors for activity recognition, given the utilization of suitable processing techniques and models.

\vspace{-2mm}
\subsection{Recognition Module}
\subsubsection{Preprocessing}  

\paragraph{Noise Removal}
The captured voltage measurements reveal the presence of high-frequency noise, necessitating the implementation of a discrete-time low-pass filter to effectively eliminate this unwanted interference. To strike a balance between mitigating the high-frequency noise and preserving the crucial components associated with human activity, we have employed a carefully chosen cutoff frequency for the filter.  This cutoff frequency is set at $5Hz$ to ensure the successful removal of undesirable noise while retaining the essential features necessary for accurate activity recognition.

\color{black}\vspace{-1mm}
\paragraph{Sequence Preparation}
In order to train the model to capture the sequential nature of human activity from the input photovoltaic measurements, we utilize a sliding window approach. This involves partitioning the photovoltaic readings into consecutive segments of fixed length. However, determining the appropriate length of the sliding window poses a challenge due to the variability in activity patterns among different individuals.
To address this issue and enhance the effectiveness of deep model training, we employ a varying window overlap technique as a form of data augmentation. By utilizing overlapping windows of varying sizes, we aim to capture a more comprehensive representation of the underlying activity sequences. This approach has been shown to improve system accuracy, as shown in Section \ref{sec:eval_window_overlap}.

Mathematically, let $L$ denote the length of the sliding window and $O$ represent the overlap size. The overlapping windows are obtained by shifting the starting position of each subsequent window by a fraction of the overlap size. This can be expressed as:
\begin{equation}
\text { Window }_i=\text { Photovoltaic }[i \times(L-O):(i \times(L-O))+L]
\end{equation}
where $\text{{Photovoltaic}}$ denotes the input photovoltaic measurements and $i$ corresponds to the window index.

By incorporating window overlap, we ensure that the model receives a diverse range of input sequences, capturing both short-term and long-term dependencies in the data. This augmentation technique contributes to the overall accuracy and robustness of the deep learning system. Further details and analysis of the impact of window overlap can be found in Section \ref{sec:eval_window_overlap}.

% \vspace{-0.5cm}

\subsubsection{Feature Extraction}
The feature extraction module plays a crucial role in enabling the classification model to focus solely on activity-related patterns, effectively disregarding noise and environmental factors.
Each captured scan represents the photocurrent readings obtained from the worn photovoltaic cells.  It is important to note that these photocurrent readings are absolute values and are highly influenced by the ambient light conditions, such as the type of light source and any changes in lighting intensity, such as sunlight or moonlight. These variations in ambient light can significantly affect the absolute values of the photocurrent readings. To address this issue, the proposed system employs a technique that calculates relative features to mitigate the interference caused by ambient light. This is achieved by computing the differences between pairs of photovoltaic cells within each sample. The rationale behind this approach is that additional ambient light typically introduces a fixed offset value to the original measurements of each cell. By taking the difference between the pairs of cells, this offset value can be eliminated, resulting in relative differential features.

Given a sample scan denoted as $V = (V_1, V_2, ... , V_m)$, the proposed system calculates the difference $\Delta V_{ab}$ between the photocurrent values of every pair of cells $V_a$ and $V_b$ using the equation $\Delta V_{ab} = V_a - V_b$.
For example, $V_{12}$ may represent the difference between the photocurrent values measured by the cells attached to the left foot $V_1$ and the right wrist  $V_2$.
A total combinations of $\binom{m}{2}$ differences are computed for each scan, resulting in a new feature vector denoted as $\Delta V = (\Delta V_{12}, 
 \Delta V_{13}, ... , \Delta V_{(m-1)m})$. 
This difference vector serves as the input relative feature vector for the activity recognition model, which will be further defined in the subsequent section.

The feature extraction method involves the subtraction of the immediate preceding time value, denoted as $V_{previous}$, from the current value, denoted as $V_{now}$. This approach calculates the temporal difference between consecutive data points, enabling the capture of the temporal dynamics inherent in the sensor readings.
This method serves as an additional technique for feature extraction, specifically focusing on the temporal aspects of the data. By subtracting the previous value from the current value, the resulting feature represents the instantaneous change in the sensor measurements. This type of feature provides valuable insights into the rate and magnitude of variations occurring over time.
In terms of filters, this approach can be categorized as a temporal differencing or temporal derivative filter. Its operation revolves around the concept of differencing adjacent data points to highlight the temporal dynamics within the signal. Through the subtraction of the previous value from the current value, this filter effectively emphasizes the temporal changes present in the data.

\begin{figure}[!t]
   % \begin{tabular}{cc}
            \vspace{-0.4cm}
            \hspace{-0.9cm}
      \begin{minipage}[t]{0.45\hsize}
%\centering
        \includegraphics[width=0.8\linewidth,height=5.2cm]{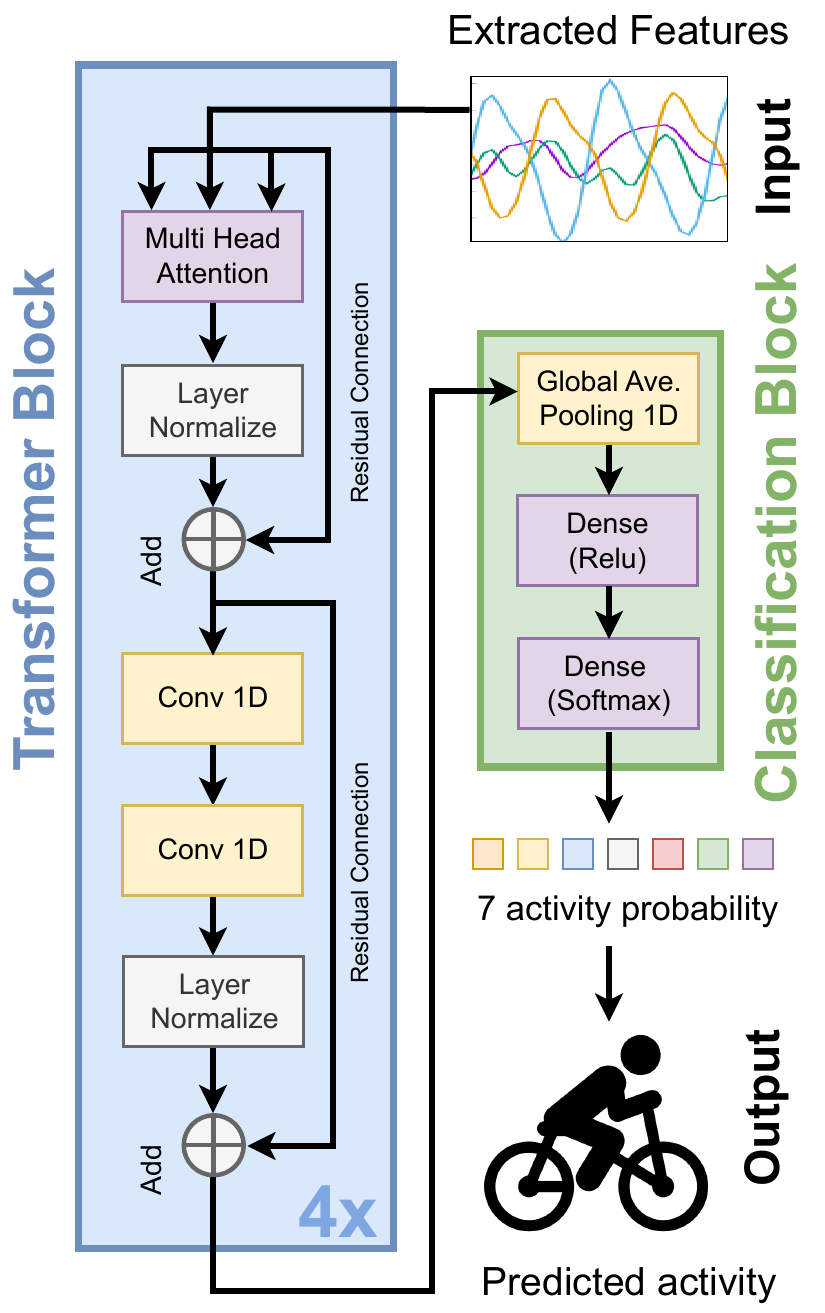} 
        \vspace{-0.45cm}
        \caption{Model arch.}
        \label{fig:model}
    \end{minipage} %&
     %\hspace{0.12\columnwidth}
         \hspace{-0.6cm}
     \begin{minipage}[t]{0.55\hsize}
    \vspace{-4.6cm}
           \captionof{table}{Default parameters of employed transformer model.}  \vspace{-0.2cm}     \centering
    \begin{tabular}{|c|c|} \hline
          \textbf{Parameter}    & \textbf{Default} \\ \hline 
        Num. of Transformer blocks    & 4  \\ \hline
        Num. of heads & 4 \\ \hline 
        Head size  & 256 \\ \hline
        %ff_dim
        Num. of output channels  & 4 \\ \hline
       Dropout rate in head  &  0.125 \\ \hline
       Dropout rate in MLP  &  0.2\\ \hline
      Num. of units of MLP    & 128  \\ \hline
    \end{tabular} \\ \vspace{1mm}

    \label{tab:transformer_param}

        \end{minipage}  \vspace{-6mm}
   % \end{tabular}
\end{figure}

\subsubsection{Activity Recognition module}

In this section, we present a  description of the multiheaded transformer network employed for activity recognition using solar cells as sensors. The transformer architecture has proven to be highly effective in various natural language processing tasks and has recently gained popularity in computer vision applications. By employing the transformer framework in our activity recognition task, we aim to leverage its ability to capture complex dependencies and relationships in the input data, leading to accurate and robust classification results.

The multiheaded transformer network consists of two main components (as shown in Figure~\ref{fig:model}): an encoder and a classifier. The encoder module captures the temporal and spatial dependencies of the extracted features, while the classifier module employs multiple attention heads to perform activity recognition based on the encoded representations.
The encoder is composed of multiple stacked transformer layers, each containing a self-attention mechanism and feed-forward neural networks. The self-attention mechanism allows the network to weigh the importance of different features based on their relevance to the current activity being performed. The feed-forward neural networks provide non-linear transformations to enhance the representation power of the model.
The input to the transformer network is a sequence  of the extracted features obtained from the subtraction of each pair of solar cells. These features reflect the variations in the photovoltaic signals captured by the cells and encode valuable information about the activities being performed. 
% The extracted features are organized in a sequence, where each element corresponds to a specific time step.

The self-attention mechanism is a pivotal component of the transformer network. It allows the model to focus on different parts of the input sequence while considering the interactions between them. This mechanism calculates attention weights for each element in the sequence, indicating the importance of that element with respect to the others. Considering these weights, the model can dynamically adjust its focus and learn the relevant context for accurate activity recognition.
To capture diverse aspects of the input sequence, the transformer network employs multiple attention heads. Each attention head independently attends to different parts of the input, allowing the model to learn different types of dependencies. These heads learn to extract relevant features from various combinations of input elements, providing a more comprehensive understanding of the input sequence.

The outputs of the encoder module are passed to the classifier, which consists of a fully connected layer followed by a softmax activation function. The fully connected layer projects the encoded representations onto the activity classes, and the softmax function normalizes the scores, producing a probability distribution over the activity classes. The class with the highest probability is then selected as the predicted activity label. We employ a  cross-entropy loss function to measure the dissimilarity between the predicted activity probabilities and the ground truth labels. The model parameters are optimized using Adam to minimize the loss function and update the model weights accordingly.

\begin{figure}[t]
        %\centering      
        \vspace{-0.4cm}
        \includegraphics[width=1\linewidth]{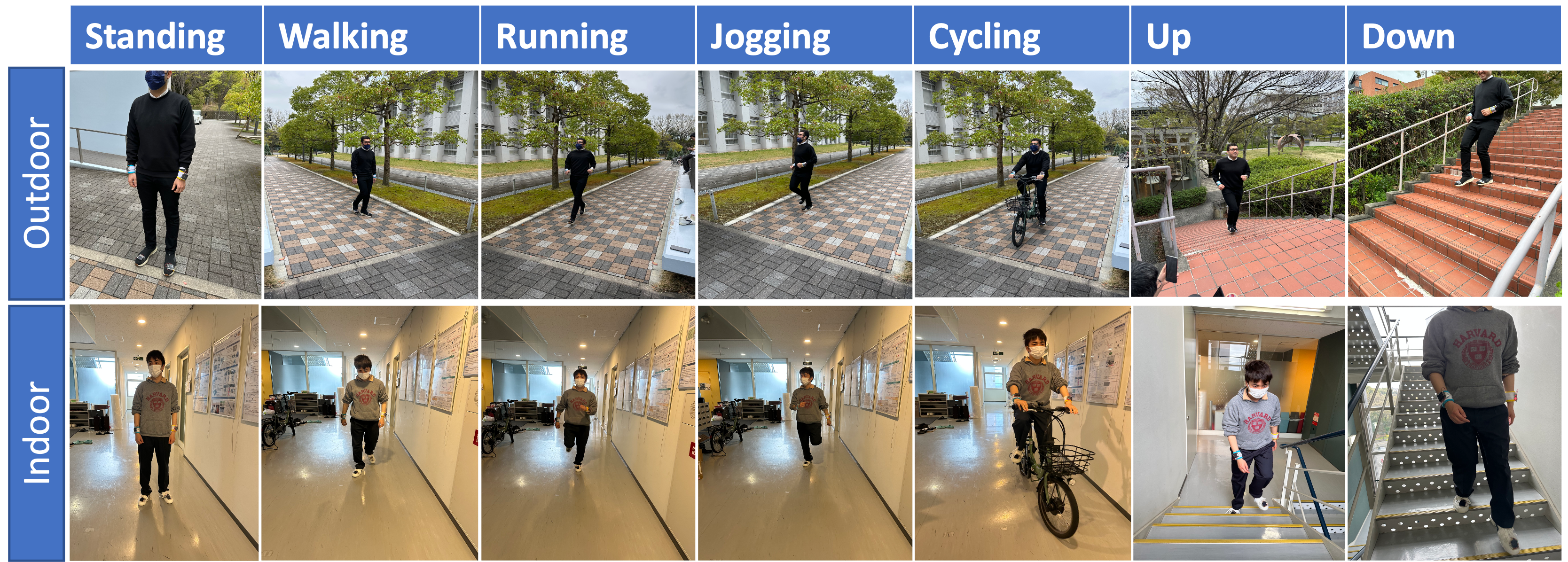} \vspace{-0.7cm} 
        \caption{Data collection of the considered activities in outdoor and indoor settings.}
        \label{fig:har}
        \vspace{-6mm}
  \end{figure}
\vspace{-1mm}
\section{Evaluation}
\subsection{Environmental Setup}

The evaluation phase involved conducting various indoor and outdoor activities (shown in Figure \ref{fig:har}), including standing, walking, jogging, running, cycling, stair climbing, and stair descending. These activities were performed by a group of six participants on separate days during different time periods, including the morning, afternoon, and evening.
During each activity, the participants wore four devices, specifically photovoltaic cells connected to micro:bits. Two devices were placed on the participants' left and right wrists, while the other two were attached to their left and right feet. These sensors record the photovoltaic values and transmit them to a server at a frequency of $23.1$ Hz.
Due to the inherent delay caused by communication and computation processes in each micro:bit, the acquisition times of the data were not strictly simultaneous. To address this issue, we employed linear interpolation to synchronize the captured values. In total, we collected 4400 samples from the participants.

To assess the performance of our system, we employed K-fold cross-validation, specifically utilizing a value of $k=5$. This approach involves dividing the dataset into five equal parts, using four parts for training and one part for testing in each iteration, while rotating the test set across all the partitions.

\vspace{-1mm}
\subsection{Parameter Evaluation}

\begin{table}[h]
    \centering      \vspace{-0.35cm}
     \caption{Default parameters for evaluation.}
     \vspace{-0.3cm}
    \begin{tabular}{|c|c|} \hline
     \textbf{Parameter}   & \textbf{Default} \\ \hline
         Sensor position  & $WWFF$ (4 sensors) \\ \hline  
         Sampling rate    &  23.1 Hz \\ \hline
         Window overlap    &  87.5\% \\ \hline
         Window size      &  1.6 sec (Time axis)  \\ \hline
         FE method        & Both (High-pass and Subtract)    \\ \hline  
     Num. of activities & 4 activities (Stand, Cycle, Walk, Run) \\ \hline
         Train/Test Data       & Outdoor, 6 subjects, 3 minutes each activity  \\ \hline  
        % Test Data        & Outdoor, 6 subjects, 3 minutes each activity  \\ \hline
         Validation       & 5-Fold Cross Validation \\ \hline
    \end{tabular}
    \label{tab:data_param}
\vspace{-0.5cm}
\end{table}

\begin{figure*}[htbp]
\vspace{-0.4cm}
    \begin{tabular}{ccc}
        \hspace{-0.8cm}
        \begin{minipage}[t]{0.33\hsize}
        %\centering
        \includegraphics[keepaspectratio, scale=0.38]{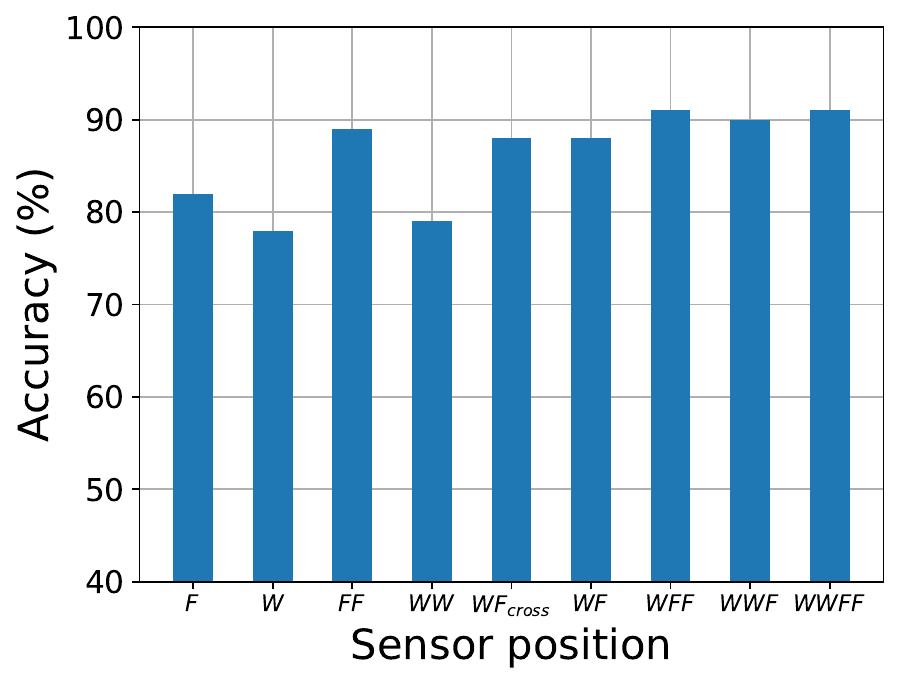}
         \vspace*{-4mm}
        \caption{Impact of sensor positions on accuracy. }
        \label{fig:senser_position}    %\vspace*{6mm}
      \end{minipage} &
          \hspace{-0.1cm}
        \begin{minipage}[t]{0.33\hsize}
        %\centering
        \includegraphics[keepaspectratio, scale=0.38]{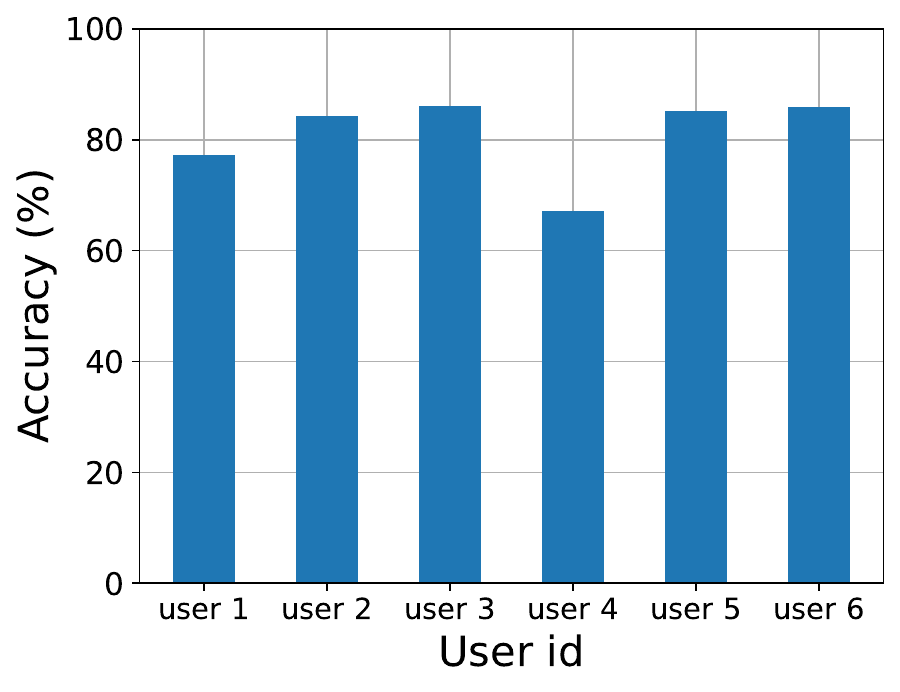}
         \vspace*{-4mm}
        \caption{Leave one user out accuracy evaluation.}
        \label{leave_one_user_out} 
      \end{minipage} &
          \hspace{0.1cm}
%       % \begin{minipage}[t]{0.33\hsize}
%       %   %\centering
%       %   \includegraphics[keepaspectratio, scale=0.38]{img/prep.pdf}
%       %   \vspace*{-4mm} 
%       %   \caption{Effect of feature extraction}
%       %   \label{fig:feature_extraction}
%       % \end{minipage} 
\hspace{-4mm}
              \begin{minipage}[t]{0.33\hsize}        \includegraphics[keepaspectratio, scale=0.37]{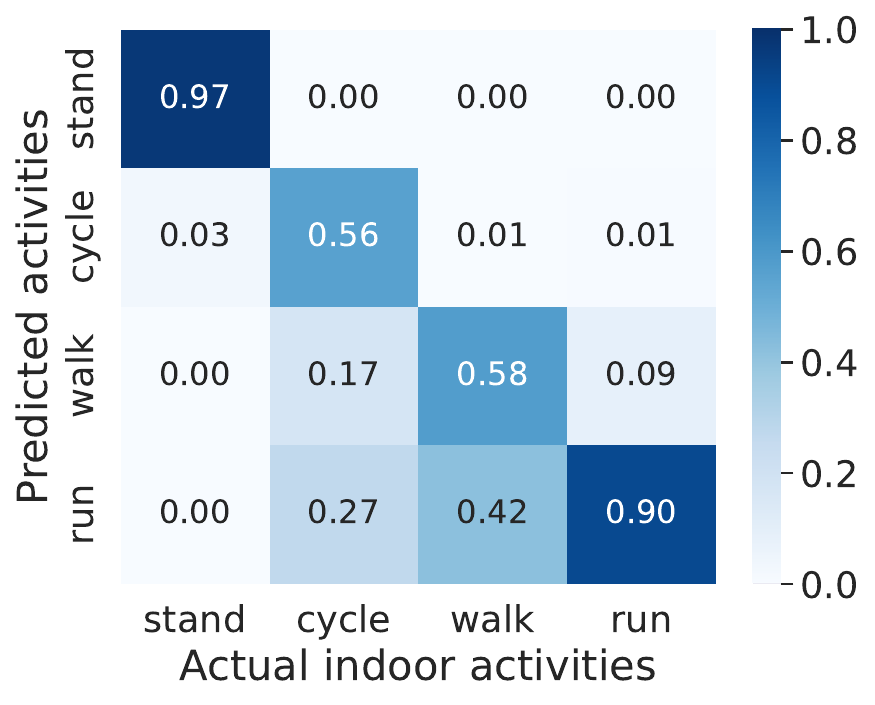}
            \vspace{-0.4cm}
        \caption{Classification result (trained outdoor and tested indoor).}
        \label{fig:cross_environment}
        \end{minipage}
  \hspace{-0.4cm}\\
%                \begin{minipage}[t]{0.33\hsize}
%         %\centering
%         \includegraphics[keepaspectratio, scale=0.38]{img/sampling.pdf}
%          \vspace*{-4mm} 
%         \caption{Effect of sampling rate.}
%         \label{fig:sampling_rate}
%       \end{minipage} &   \hspace{-0.1cm}
%               \begin{minipage}[t]{0.33\hsize}
%         %\centering
%         \includegraphics[keepaspectratio, scale=0.38]{img/window.pdf}
%          \vspace*{-4mm}
%         \caption{Effect of window size. }
%         \label{fig:window_size}
      % \end{minipage}  &   \hspace{-0.1cm}
                   \hspace{-6mm} 
   \begin{minipage}[t]{0.33\hsize}
        %\centering 
        \includegraphics[keepaspectratio, scale=0.37]{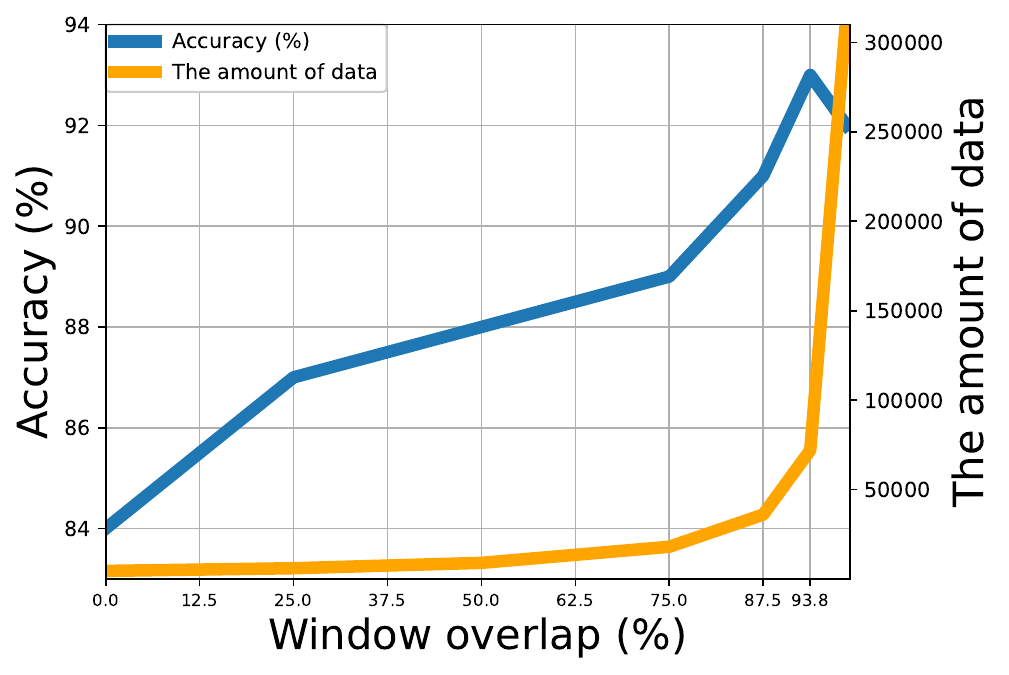}
         \vspace*{-8mm}
        \caption{Impact of overlap rate on accuracy.}
        \label{fig:overlap_rate} 
      \end{minipage}  \vspace*{6mm} & \hspace{2mm}
      \begin{minipage}[t]{0.33\hsize}
            \includegraphics[keepaspectratio, scale=0.4]{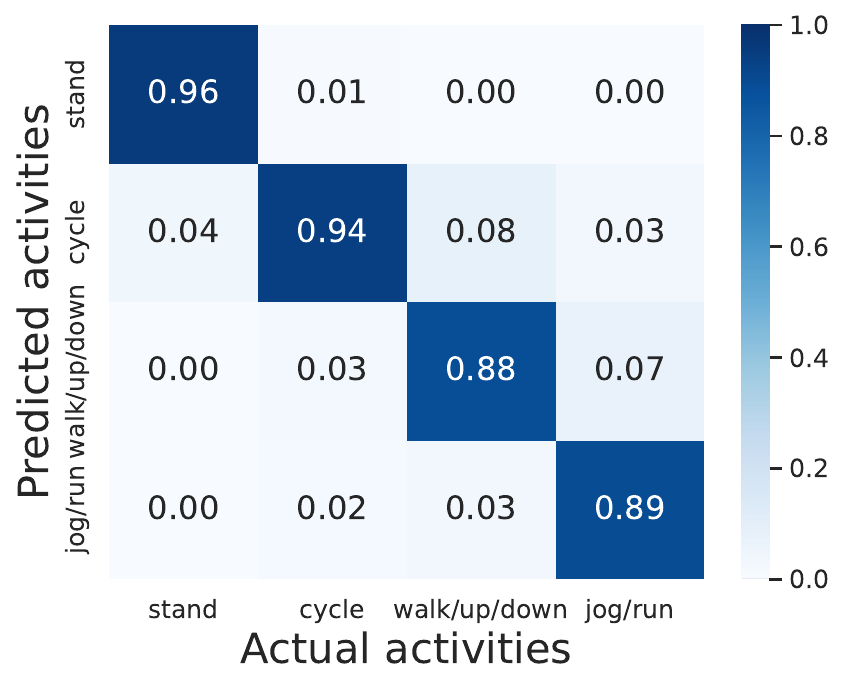}
            \vspace{-0.4cm}
            \caption{ Classification result (4 activities with default settings).}
            \label{fig:default_4act}
        \end{minipage} & \hspace{-2mm}
        \begin{minipage}[t]{0.33\hsize}
            \includegraphics[keepaspectratio, scale=0.4]{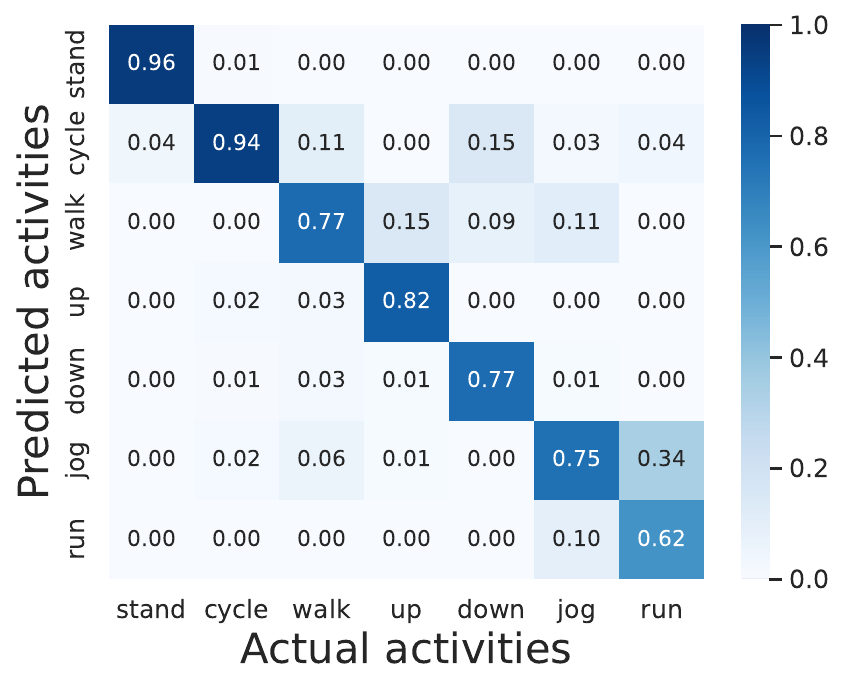} 
            \vspace{-0.4cm}
            \caption{ Classification result (7 activities with default settings).}            
            \label{fig:default_7act}
        \end{minipage}        
    \end{tabular}
    \vspace{-1.1cm}
    
\end{figure*}

In this section, we analyze the impact of several system parameters, including sensor position, sampling rate, percentage of window overlapping, and window size, on the overall performance. The default values for these parameters can be found in Table \ref{tab:data_param}.

 We have fixed the internal parameters of the transformer for the purpose of impartial experimentation as shown in Table \ref{tab:transformer_param}.

\subsubsection{Sensor Position}
The impact of varying the sensor position on the activity recognition accuracy of the system is evaluated in this section, as depicted in Figure \ref{fig:senser_position}.
The experiment involves a maximum of four sensors in total, with two sensors attached to the wrist and two to the foot.
One or two sensors are positioned either on foot (denoted as $F$ or $FF$), the wrist (denoted as $W$ or $WW$), both (denoted as $WF$ or $WF_{cross}$ for the opposite limbs), or all (denoted as $WWFF$). 
Upon analyzing the findings, it becomes evident that foot-based measurements play a more crucial role in accurately classifying the  activities compared to wrist-based measurements. This indicates that the information obtained from sensors positioned on foot carries greater discriminative power and relevance for activity recognition.
The observed superiority of foot-based features can be attributed to the inherent nature of the activities under consideration. Certain movements and patterns associated with the predefined activities, such as walking or running, are predominantly reflected in the motion and positioning of the lower extremities. Consequently, capturing data from the foot sensors allows for a more precise and representative characterization of these activities, leading to improved recognition accuracy.
On the other hand, while wrist-based sensors still contribute to activity recognition, their relative importance appears to be lower in comparison. This could be due to factors such as reduced motion range or the influence of upper body movements being less pronounced in the specific activities considered in this study.
\vspace{-0.3mm}

\color{black}
\subsubsection{Percentage of Window Overlapping} \label{sec:eval_window_overlap}
In this section, we study the percentage of window overlap on the system overall performance. The overlap percentage of the sliding window $\frac{O}{L} \times 100$ was chosen from a range of values, including 0\%, 25\%, 50\%, 62.5\%, 75\%, 87.5\%, 93.75\%, and 99\%. This selection aimed to evaluate the impact of different overlap configurations on the accuracy of the system. Notably, through rigorous experimentation and analysis, it was observed that an overlap percentage of 93.75\% yielded the highest accuracy, as depicted in Figure \ref{fig:overlap_rate}. This can be justified as at this level the system strikes a balance between capturing sufficient contextual information within each window while minimizing redundancy and potential data leakage between consecutive windows. This configuration enables the model to effectively learn the underlying sequential patterns present in the photovoltaic measurements, enhancing its predictive capabilities.

\color{black} \vspace{-1mm}
\subsection{Performance Evaluation}

\subsubsection{Default System Performance}

The default system performance is depicted in Figure \ref{fig:default_4act} and Figure \ref{fig:default_7act}, presenting the confusion matrices for the four and seven activity categories, respectively. These results are based on the settings specified in Table \ref{tab:data_param} and 5-fold cross-validation. The results highlight the remarkable accuracy achieved by the system, reaching 91.7\% for the four activity categories and 80.4\% for the seven activity categories. This outstanding performance can be attributed to the effective feature extraction methodology employed and the robust learning capability of the transformer model. Together, these factors contribute to the system's accurate and efficient recognition of activities.

\color{black} 
\subsubsection{Leave-One-User-Out Performance}

In this section, we assess the performance of the proposed activity recognition system using a leave-one-user-out evaluation approach. The system is trained on data from five users and tested on the data from the remaining user. Figure \ref{leave_one_user_out} illustrates the results, which demonstrate the system's strong generalization capabilities for unseen users, with a consistent  accuracy of approximately 80\%.
This can be attributed to the inherent generalization abilities of the adopted transformer model. Transformers are known for their capacity to capture long-range dependencies and learn representations that are transferable across different instances or individuals. In the context of activity recognition, this translates to the model's ability to extract relevant features and patterns from the training data that are applicable to new, unseen users during the testing phase.

\color{black} 
\subsubsection{Cross-Environment Performance}

The evaluation of the model's performance in a cross-environment scenario is depicted in Figure \ref{fig:cross_environment}. In this evaluation, the model is trained on photovoltaic data of activities conducted in outdoor environments. Subsequently, the inference is performed on test data of the same activities conducted indoors.
One notable challenge in cross-environment activity recognition arises from the significant difference in average light intensity between indoor and outdoor settings. Consequently, the datasets obtained from these environments exhibit distinct amplitude characteristics. 
However, the proposed system demonstrated a commendable level of performance and robustness even when confronted with an environment that starkly differed from the training setting. It achieved an accuracy exceeding 75\%, thereby showcasing its ability to generalize well across varying environmental conditions.
This can be attributed to the feature extraction method employed, which focuses on extracting relative features rather than relying solely on the absolute amplitude of the photovoltaic data. By adopting this approach, the recognition model becomes capable of learning the intrinsic activity patterns regardless of the specific lighting scenario. 
\vspace{-1mm}
\subsection{Discussion and Limitations}
 \setlength{\leftmargini}{10pt} 
\begin{itemize} 
\item \textbf{Energy-related concerns:}
While the proposed system has undergone rigorous testing both indoors and outdoors, encompassing a realistic range of conditions, its operational requirements necessitate a minimal amount of energy for communication and computing. Achieving this objective becomes unattainable under dim light conditions. 
 Consequently, we have expanded our design approach by integrating a compact battery component into the system. This component opportunistically harnesses available light for charging purposes, even during periods of sensor idleness. Notably, this augmentation not only facilitates the stabilization of energy requirements for the computing and communication modules but also serves the purpose of mitigating the challenges posed by inadequate lighting conditions.

\item \textbf{Sensing-related issues:}
  % is highly dependent on the brightness of the
  %   environment because it uses a solar system.
  %   For example, it would be good to add a discussion and future policy on whether the
  %   system can be used in dimly lit conditions in the evening or under streetlights at
  %   night. For
  %   example, what happens if the person has a sleeve that covers the cell? what
  %   happens at different times of the day when the sunlight and angle of incidence is
  %   changing? what is the robustness of the system?
  %   - The study is limited not comprehensive as it doesn't study the main issue of the
  %   sensor which are lightning conditions (i.e. hours of day, cloudy, different indoor
  %   lights)
The proposed system may encounter arduous scenarios in sensing that have a substantial impact on performance. One such example is the inadvertent obstruction of light before it reaches one or more solar cells.
Nonetheless, this issue can be addressed by training the model to handle such occurrences. In essence, the model can be trained using partial inputs by employing random input dropout. This approach helps mitigate the interdependence of input features, enabling the transformer model to achieve greater generalization capabilities.

   % \item \textbf{Future work:} It is part of our future research agenda to quantitatively assess the system's performance in increasingly demanding scenarios.
    % different situations and lightning conditions (i.e. hours of day, cloudy, different indoor lights, etc.) with larger group of participants.
    
\end{itemize}
\color{black} \vspace{-3mm}
\section{Conclusion}

In this paper, we introduced an energy-free activity recognition system leveraging photovoltaic cells as a sensor. By utilizing photovoltaic cells attached to the wrist and foot, the system seamlessly realizes  activity recognition and energy harvesting functionalities. The system is designed to accurately recognize activities across diverse subjects, time variations, and lighting conditions. The evaluation conducted in realistic indoor and outdoor environments demonstrates the system's robustness, achieving an accuracy of 91.7\%. The proposed system has the potential to revolutionize wearable technology and IoT devices by enabling self-sustaining and uninterrupted activity recognition.

% We have successfully classified seven categories of human activities with high accuracy using a lightweight and small solar panel as a wearable sensor.
% The proposed architecture shows a certain degree of robustness even in environments that differ significantly from the training environment, demonstrating that commercially available solar cells have the potential to become a new energy-free sensing method.

% \begin{figure*}[t]
%     \begin{tabular}{cc}
%         \begin{minipage}[t]{0.5\hsize}
%             \includegraphics[keepaspectratio, scale=0.55]{img/4act.pdf}
%             \caption{One user out performance in 4 activities}
%             \label{fig: one_user_out_4act}
%         \end{minipage} 
%         &
%         \begin{minipage}[t]{0.5\hsize}
%             \includegraphics[keepaspectratio, scale=0.55]{img/7act.pdf}
%             \caption{One user out performance in 7 activities}          
%             \label{fig:one_user_out_7act}
%         \end{minipage} \\
%     \end{tabular}
% \end{figure*}

% \begin{figure}[ht]
%         %\centering
%         \includegraphics[width=0.8\linewidth]{img/sensor_position.pdf}
%         \caption{
%         \color{blue}
%         The effect of sensor position. "W" and "F" correspond to one wrist sensor and foot sensor, respectively, and "WF\_cross" is two sensors in the opposite of foot and hand. "WF" means two sensors in the half of the body on the same side.}
%         \label{fig:sensors}
% \end{figure}

\color{black}

\vspace{1mm}
% \begin{acks}
\textbf{ACKNOWLEDGMENTS}
This work was supported in part by JSPS KAKENHI Grant number 22K12011 and 
19H05665, Japan.
% \end{acks}

\vspace{-1mm}
\bibliographystyle{ACM-Reference-Format}
\bibliography{_ref}

\end{document}